\documentclass[11pt]{article}
\usepackage{amssymb,amsmath,cite,geometry,graphicx,url}
\catcode`\@=11

\@addtoreset{equation}{section}
\def\theequation{\arabic{section}.\arabic{equation}}
     
\catcode`\@=11
\def\thesection{\arabic{section}.}

\def\appendix{\setcounter{section}{0}
        \def\thesection{Appendix.}
        \def\theequation{\Alph{section}.\arabic{equation}}}
\def\section{\@startsection{section}{1}{\z@}{3.5ex plus 1ex minus
   .2ex}{2.3ex plus .2ex}{\large\bf}}
     
\long\def\@makefntext#1{\parindent 0cm\noindent
\hbox to 1em{\hss$^{\@thefnmark}$}#1}

\newcommand{\captionfonts}{\small}
\makeatletter  
\long\def\@makecaption#1#2{%
  \vskip\abovecaptionskip
  \sbox\@tempboxa{{\captionfonts #1: #2}}%
  \ifdim \wd\@tempboxa >\hsize
    {\captionfonts #1: #2\par}
  \else
    \hbox to\hsize{\hfil\box\@tempboxa\hfil}%
  \fi
  \vskip\belowcaptionskip}
\makeatother   

\begin{document}
\begin{titlepage}
\vspace{.5in}
\begin{flushright}
September 2018\\  
revised July 2019
\end{flushright}
\vspace{.5in}
\begin{center}
{\Large\bf
 Hiding the cosmological constant}\\  
\vspace{.4in}
{S.~C{\sc arlip}\footnote{\it email: carlip@physics.ucdavis.edu}\\
       {\small\it Department of Physics}\\
       {\small\it University of California}\\
       {\small\it Davis, CA 95616}\\{\small\it USA}}
\end{center}

\vspace{.5in}
\begin{center}
{\large\bf Abstract}
\end{center}
\begin{center}
\begin{minipage}{4.45in}
{\small
Perhaps standard effective field theory arguments are right, and vacuum fluctuations really 
\emph{do} generate a huge cosmological constant.  I show that if one does not assume 
homogeneity and an arrow of time at the Planck scale, a very large class of general relativistic
initial data exhibit expansions, shears, and curvatures that are enormous at small 
 scales, but quickly average to zero macroscopically.  
 Subsequent evolution is more complex, but I argue that quantum fluctuations may 
 preserve these properties.  The resulting picture is a version of Wheeler's ``spacetime 
 foam,'' in which the cosmological constant produces high curvature at the Planck scale 
 but is nearly invisible at observable scales. 
}
\end{minipage}
\end{center}
\end{titlepage}
\addtocounter{footnote}{-1}

\section{The cosmological constant problem}

Quantum fluctuations of the vacuum are expected to generate a very high energy 
density, which should manifest itself as an enormous cosmological constant.  We don't
know how to calculate this quantity exactly, and it remains possible that it is suppressed, 
exponentially \cite{Holland} or otherwise \cite{Martin}.  But standard effective field theory 
arguments predict a value $\Lambda\sim \pm 1/\ell^2$, where the cut-off length $\ell$ is usually 
taken to be the Planck length $\ell_P$ \cite{EFT,Carroll}. The sign of $\Lambda$ depends 
on the exact particle content of the universe---bosons and fermions contribute with opposite 
signs---but unless a remarkable cancellation occurs, the predicted value is huge.

We do, in fact, observe an accelerated expansion of the universe that could be due to 
a cosmological constant.  But a Planck-scale cosmological constant is some 120 orders 
of magnitude too large, making it what has been called ``the worst theoretical prediction 
in the history of physics'' \cite{Hobson}.  It is widely assumed that $\Lambda$ 
must either be canceled by incredibly precise fine tuning or eliminated by some other form 
of special pleading---anthropic selection \cite{Weinberg}, nonlocal modifications of the 
gravitational action \cite{Kaloper}, or the like.  The problem is made especially intractable by 
the mixing of scales: $\Lambda$ is generated near the Planck scale, but observed at 
cosmological scales.

Here I propose a simple but radical alternative.  Perhaps our universe really \emph{does} 
have a cosmological constant of order $1/\ell^2$, with $\ell$ possibly as small as  $\ell_P$.
In a homogeneous universe this would  be immediately ruled out by observation.  But if 
$\Lambda$ is generated by Planck scale fluctuations, there is no reason to expect homogeneity 
at that scale.  This notion was anticipated by Wheeler \cite{Wheeler}, who called the
resulting picture ``spacetime foam.''  
Note that I am not considering fluctuations of the cosmological constant itself, though
they may also matter \cite{Unruh}.  Rather, I am proposing that the \emph{effects} of
the cosmological constant may fluctuate in a way that averages to near zero.  

For a space with a positive definite metric, it is easy to imagine high curvature at small 
scales averaging to zero macroscopically.    For a spacetime, though, a cosmological 
constant would seem to entail exponential expansion (if the anisotropy is not too big 
\cite{Wald}), and it is not clear how such behavior can be averaged away.  But this picture
is too simple: a cosmological constant can produce either expansion or contraction, and as 
we shall see, this behavior can vary at the Planck scale. Over a larger region, a large
$\Lambda$ may thus be consistent with small average expansion.
  
In what follows, I make this idea more concrete.  Starting with the initial value
formulation of general relativity with an arbitrary cosmological constant, I show that 
a very large class of initial data has a local Hubble constant that is huge at the Planck scale 
but tiny macroscopically.  For an infinite subset of data, the macroscopic spatial curvature is
also very small, and has a vanishing first time derivative.  A ``macroscopic''  region
here need not be very large: a cubic centimeter already contains some $10^{100}$ 
Planck-size regions.

An initial value formulation is not enough---one must also show that these features
are preserved dynamically.  Higher order time derivatives depend on finer details, and are
harder to analyze.  If the initial inhomogeneities are generated by quantum fluctuations, 
though, I argue that these fluctuations should also preserve the crucial properties that 
camouflage the cosmological constant.

These arguments do not provide a complete answer to the cosmological constant problem.  
They do not, for example, explain the apparent existence of a very small $\Lambda$
at macroscopic scales.  More generally, one would have to show that long wavelength
excitations on this foam-like background obey a macroscopic version of the Einstein field 
equations, a form of the much-studied but unresolved ``averaging problem'' \cite{Clarkson}.  
But the results here suggest, at least, that we may have been looking for answers
at the wrong scales.

\section{The initial value formulation}

Let $\Sigma$ be a compact three-dimensional manifold, interpreted as a Cauchy surface
of a spacetime.  The initial data for general relativity on $\Sigma$ consist of a spatial metric 
$g_{ij}$ and an extrinsic curvature $K^i{}_j$.  These are not arbitrary, but must satisfy a set of 
constraints.  If the contribution of matter is negligible compared to the cosmological constant,
these are
\begin{subequations}
\begin{align}
&R + K^2 - K^i{}_jK^j{}_i - 2\Lambda = 0 \, ,\label{a1a}\\
&D_i(K^i{}_j - \delta^i_jK) = 0 \, ,\label{a1b}
\end{align}
\end{subequations}
where $R$ is the scalar curvature of the metric $g_{ij}$, $D_i$ is the covariant derivative compatible
with that metric, and $K=K^i{}_i$.  This  is the formalism that translates most natural 
into canonical quantum theory; the constraint  (\ref{a1a}) becomes the Wheeler-DeWitt equation,
while (\ref{a1b}) imposes spatial diffeomorphism invariance.
It is sometimes useful to split off the trace of the extrinsic curvature, writing
\begin{equation}
K^i{}_j = \sigma^i{}_j + \frac{1}{3}K\delta^i_j  \, .
\label{a3}
\end{equation}
$K$ is the expansion---it is the local Hubble constant, the logarithmic derivative of the 
volume element---while $\sigma^i{}_j$ is the shear tensor.  The shear scalar is defined as
$\sigma^2=\frac{1}{2}\sigma^i{}_j\sigma^j{}_i$.

The dynamical evolution of this data is described by the equations
\begin{subequations}
\begin{align}
&\mathcal{L}_n g_{ij} = 2 g_{ik}K^k{}_j \, , \label{a2a}\\
&\mathcal{L}_n K^i{}_j = -R^i{}_j - KK^i{}_j + \Lambda\delta^i{}_j  + \frac{D^iD_jN}{N} \, ,  \label{a2b}
\end{align}
\end{subequations}
where $\mathcal{L}_n$ is the Lie derivative along the unit normal to $\Sigma$, essentially a
covariant time derivative, and $N$ is the lapse function, which determines the  
position-dependent separation of successive time slices.  (For simplicity, I have taken the shift
vector to be zero.)  $N$ must be positive, but it is otherwise arbitrary, making the evolution
appear nonunique; but solutions with different choices of $N$ are related by diffeomorphisms,  
and are thus physically equivalent.

The evolution equations (\ref{a2a})--(\ref{a2b}) and the constraints (\ref{a1a})--(\ref{a1b}) have
rather different status in quantum gravity.  Assuming a gravitational version of Ehrenfest's theorem, 
the evolution equations should hold for averages, but observed values of geometric quantities 
will be subject to quantum fluctuations, presumably of order one at the Planck scale.  The
constraints are different: while their precise form may be modified by quantum
effects,  some version of the constraints is likely to hold \emph{exactly}.    In an operator formalism, 
for instance, the Wheeler-DeWitt equation is the statement that the constraints exactly annihilate 
physical states \cite{WdW}, while in typical path integral approaches, only configurations that satisfy the 
constraints appear in the sum over histories (though with some ambiguity \cite{Teitelboim}).  The constraints
thus capture the quantum structure at the Planck scale in a way the evolution equations do not.  

We will now need two properties of the initial value formulation:
\begin{enumerate}
\item The equations are time reversal invariant: if $(g,K)$ is allowed initial data for a manifold
$\Sigma$, so is $(g,-K)$.
\item Two manifolds $\Sigma_1$ and $\Sigma_2$ with initial data $(g_1,K_1)$ and $(g_2,K_2)$
can be ``glued'' to form a manifold $\Sigma_1\#\Sigma_2$ for which the initial data is unchanged 
outside arbitrarily small neighborhoods of the points where the gluing is performed 
\cite{Chrusciel,Chruscielb}.

More precisely, $\Sigma_1\#\Sigma_2$ is topologically  the connected sum of $\Sigma_1$ and
$\Sigma_2$, formed by cutting balls out of each manifold and identifying the 
boundaries.  Geometrically, pick open sets $U_1\subset\Sigma_1$ and $U_2\subset\Sigma_2$,
restricted only by the generic condition that the initial data is ``not too symmetric,'' in the
sense that the domains of dependence of $U_1$ and $U_2$ contain no Killing vectors.
Pick points $p_1\in U_1$ and $p_2\in U_2$, cut geodesic balls $B_1$ and $B_2$ of arbitrarily small 
radius around each, and join the boundaries.  Then $\Sigma_1\#\Sigma_2$ admits initial data
that exactly coincides with the original data outside $U_1\cup U_2$ and is close to the original data, 
in a suitable norm, inside $U_1\cup U_2$ but outside $B_1\cup B_2$.
\end{enumerate}

Now, as a preliminary construction,  
pick a three-manifold $\Sigma$ with a fixed 
open set $U$ and a point $p\in U$, and specify initial data $(g,K)$.   Let $\bar\Sigma$ be an identical
copy of $\Sigma$, but with initial data $(g,-K)$.  The two manifolds can be glued symmetrically at $p$ 
(see section 5 of \cite{Chruscielb}) to form a connected sum ${\widetilde\Sigma} = \Sigma\#{\bar\Sigma}$.
By symmetry, $\widetilde\Sigma$ will have a reflection isometry, under which  $(g,K)\rightarrow(g,-K)$.  
While the definition of an averaged tensor is ambiguous \cite{Clarkson},  any average that respects this 
symmetry will clearly give $\langle K^i{}_j\rangle = 0$.

Next, much more generally, consider a large collection of manifolds $\Sigma_1,\Sigma_2,\dots,
\Sigma_N$, each with its own initial data $(g_\alpha,K_\alpha)$.  Form the glued manifold
\begin{equation}
{\widetilde\Sigma} = \Sigma_1\#\Sigma_2\#\dots\#\Sigma_N \, .
\label{z1}
\end{equation}
As long as we do not assume a microscopic arrow of time, the data $(g,K)$ and $(g,-K)$ for any 
particular $\Sigma_\alpha$ will be equally likely.  Thus, again, any sensible 
average over a large enough number of components should give $\langle K^i{}_j\rangle \sim 0$.  Exactly 
how fast the average will go to zero depends on the number and distribution of manifolds and initial 
data sets, but even a cubic centimeter contains some $10^{100}$ Planck-size regions.

These results imply that $\langle \mathcal{L}_n\sqrt{g} \rangle= 0$
and $\langle\sigma^i{}_j\rangle = 0$.  It is also easy to check that $\langle\mathcal{L}_nR\rangle = 0$.  To 
first order, the averaged spatial geometry is thus stationary.  To match our universe, we 
would also like the average spatial curvature to be small.  For the initial data, the only restriction comes 
from averaging the constraint (\ref{a1a}):
\begin{equation}
\langle R\rangle =   2\langle\sigma^2\rangle + 2\Lambda - \frac{2}{3}\langle K^2\rangle \, .
\label{a4}
\end{equation}
It is thus evident that if $\langle K^2\rangle$ is large (for positive $\Lambda$) or $\langle\sigma^2\rangle$ 
is large (for negative $\Lambda$), the cosmological constant can be ``absorbed'' in fluctuations of 
extrinsic curvature.

Let me stress that I am not starting with a spacetime and searching for a special hypersurface on which 
$\langle K^i{}_j\rangle = 0$.  That would be an artificial procedure, and there would be no
reason to expect such a hypersurface to be physically interesting.  Rather, I am taking an arbitrary 
hypersurface and giving it initial data chosen randomly from a large collection.  Further requirements
may be added to make this data ``nice,'' but as long as these allow Planck-scale inhomogeneity and 
do not pick out a microscopic arrow of time, the conclusions should not change.

This construction allows ${\widetilde\Sigma}$ to have an arbitrarily complicated topology.  Indeed, any orientable
compact three-manifold has a unique decomposition as a connected sum of ``prime'' manifolds \cite{Milnor,top}.
But ${\widetilde\Sigma}$ may also be topologically trivial: if each $\Sigma_\alpha$ is a three-sphere, the connected
sum is also a three-sphere.  We can thus reach a large set of initial data by starting with any initial values, 
cutting out a collection of Planck-size balls, changing the data on the balls, and gluing them back.  The
geometry of the ``necks'' between components is rather special, though, and it
is an open question how much of the space of initial data can be reached this way.

For the special case of local spherical symmetry, the construction can be made  
explicit \cite{Witt}.  This case is ``too symmetric'' to meet the genericity condition for the gluing theorem,
but it is possible to construct exact initial data on a space with topology $S^2\times S^1$ 
made up of alternating expanding and contracting shells with an expansion that averages to zero \cite{Carlipz}.
 
\section{Evolution \label{evo}}

We have established that on an initial hypersurface, a large class of initial data can exhibit small 
average expansion, hiding the macroscopic 
effect of a cosmological constant.  But is this feature preserved in time?
This is a hard question, whose answer almost certainly requires a better 
understanding of quantum gravity.  In particular, the evolution equations (\ref{a2a})--(\ref{a2b}) are
classical approximations, which do not include the quantum fluctuations that presumably create
the complex microscopic structure we are interested in.  Naively, we might have two expectations:

$\bullet$ Expanding regions grow in time, while contracting regions shrink, so if $\Lambda>0$ 
the expanding regions should eventually dominate in a volume average, although 
this may take an arbitrarily long time \cite{Burkhart}.   (If $\Lambda<0$, expanding regions will
recollapse, so this is less of an issue \cite{Unruhb}.)
 
$\bullet$ But nothing in this construction picks out a preferred initial time, so if the ``foamy''
structure is generated by quantum fluctuations, it should replicate itself: expanding regions 
should themselves fill up with new curvature fluctuations.

Without a better understanding of how (or whether) quantum fluctuations generate spacetime foam, it is
unlikely that we can fully resolve this question.  Still, we can look for hints from what we \emph{do}
know of the evolution.

\subsection{Classical evolution}
 
Let us first ask whether the classical evolution (\ref{a2a})--(\ref{a2b}) can preserve
the averaged structure.  This is similar to Buchert's question, in a somewhat different context 
\cite{Buchertb}, of whether a non-equilibrium ``cosmic equation of state'' can lead to a stationary 
averaged configuration.  

This is at least a well posed question, although still a difficult one.
First, we can only hope to learn about short-time evolution.  The initial data described here typically 
develop singularities, with minimal spheres in the the connecting ``necks'' forming
trapped surfaces that behave like black hole horizons \cite{Burkhart,Burkhartb}.  Independent
computations also indicate that Planck-scale fluctuations can disrupt the causal structure of 
spacetime \cite{Pitelli}.  It is generally  assumed that quantum gravity will resolve such singularities, 
but classically they signal a breakdown of evolution.
 
Second, there are well known ambiguities in defining time derivatives of averages.  To determine the 
derivative of an average $\langle\, \vcenter{\hbox{\tiny$\bullet$}}\, \rangle$ over a region $\mathcal{U}$,
we must specify how $\mathcal{U}$ changes in time.  If $\mathcal{U}$ is fixed in terms of some set 
of coordinates, the result will not be invariant; if it is defined in terms of geometric quantities, 
it will typically be time-dependent.  Further, averages are often (although not always \cite{Green}) 
defined in term of integrals with a dynamical integration measure, providing an added source of
time dependence \cite{Buchert}.   

Third, even if we know what we mean by ``average,'' it's not so clear what we mean by ``time.''  
The splitting of spacetime into space and time is not unique.  In the present formalism, this is 
reflected in the arbitrary choice of lapse function $N$.  In principle, the physics can be captured
by diffeomorphism-invariant, lapse-independent observables, but these are necessarily nonlocal 
\cite{Torre}, and  are poorly understood.  In practice, we usually refer instead to a ``preferred''
choice of lapse.  For an FLRW cosmology with homogeneous and isotropic initial data, for instance,
only a tiny class of lapse functions preserve these characteristics; the usual claim of homogeneity and 
isotropy is secretly a statement about the existence of these special lapse functions.  

In view of these problems, and taking inspiration from the FLRW example, we can ask whether 
 \emph{any} choice of lapse function preserves the averaged properties of our initial data.
Let us start with the condition $\langle K\rangle=0$.  Following \cite{Buchert}, define spatial averages as 
volume integrals,
\begin{equation}
\langle X\rangle_{\mathcal{U}} = \frac{1}{V_{\mathcal{U}}}\int_{\mathcal{U}}  X\sqrt{g}\,d^3x \quad\text{with} 
   \quad V_{\mathcal{U}} = \int_{\mathcal{U}} \sqrt{g}\,d^3x \, ,
\label{bx1}
\end{equation}
where the region $\mathcal{U}$ is defined in some time-independent way.  Then, from (\ref{a2a})--(\ref{a2b}),
\begin{multline}
\frac{d\ }{dt}\langle K\rangle =  \frac{1}{V_{\mathcal{U}}}\int_{\mathcal{U}}  N\mathcal{L}_n(K\sqrt{g})\,d^3x \\
   = \frac{1}{V_{\mathcal{U}}}\int_{\mathcal{U}} N(-R + 3\Lambda)\sqrt{g}\,d^3x 
   =  \frac{1}{V_{\mathcal{U}}}\int_{\mathcal{U}} N\left(\Lambda + \frac{2}{3}K^2 - 2\sigma^2\right)\sqrt{g}\,d^3x \, .
\label{bx2}
\end{multline}

If we choose a uniform time-slicing, $N=1$, this becomes
$d\langle K\rangle/dt = - \langle R\rangle + 3\Lambda$,
which is essentially the second Friedmann equation for averaged quantities.  
But given the ``foamy'' nature of the initial geometry, there is no reason to choose a constant lapse function
at the Planck scale.  As long as the integrand in (\ref{bx2})  doesn't have a definite sign---that is, 
as long as the shear (for positive $\Lambda$) or expansion (for negative $\Lambda$) is large in some 
regions---there will be an infinite number of choices of $N$ for which the right-hand side of (\ref{bx2}) 
vanishes.  For a topologically complicated manifold, in particular, the curvature of the prime factors
is typically negative, while that of the connecting ``necks'' is large and positive, so cancellation should
not be hard to achieve.

We can further choose $N$ to be invariant under $(g,K)\rightarrow(g,-K)$, which will
guarantee that any integrand with an odd power of $K$ will also average to zero. 
One might worry that our conditions could force the average curvature to be large.  
But while (\ref{bx2}) implies $\langle NR\rangle \sim \Lambda\langle N\rangle$, for a 
lapse function with Planck scale structure it can still be true that 
$|\langle R\rangle|\ll|\langle NR\rangle/\langle N\rangle| \sim\Lambda$.

The second derivative is also simple:
\begin{equation}
\frac{d^2\ }{dt^2}\langle K\rangle =  \frac{1}{V_{\mathcal{U}}}\int_{\mathcal{U}} \left[
   ({\dot N} + NK)\left(\Lambda + \frac{2}{3}K^2 - 2\sigma^2\right) + 2N^2 K^{ij}R_{ij}\right] \sqrt{g}\,d^3x \, .
\label{bx3}
\end{equation}
The last term contains an odd power of $K$, and goes to zero for a large enough region $\mathcal{U}$.  
The first term has exactly the same form as  (\ref{bx2}), and $\dot N$ can be specified independently, 
so if the first derivative can be made to vanish, the second derivative can as well. 

Higher derivatives are more complicated.  $\mathcal{L}_n^3K$, for instance, contains derivative terms 
like $K\Delta K$ and higher order correlations like $\langle K^4\rangle - \langle K^2\rangle^2$, which probe 
structure at shorter distances.  But each new derivative of $\langle K\rangle$ also comes with a new time 
derivative of $N$, which can be specified independently.  Hence there is thus no obvious obstruction to 
choosing a time-slicing for which all of the time derivatives of $\langle K\rangle$ vanish.

This is a strong claim, implying that despite the presence of very high curvature at the Planck
scale, there should exist---at least for short times---a foliation of spacetime by slices of vanishing 
average expansion.  Of course, such a foliation would itself vary rapidly at the Planck scale, but 
given the ``foamy'' structure of the three-geometry, that should come as no surprise.
Whether one can simultaneously choose a lapse function for which $\langle R\rangle$ remains small
is a more difficult question, requiring future work.

\subsection{Quantum evolution}

In the absence of a full quantum theory of gravity, much less can be said about quantum
evolution.  As noted earlier, though, in many approaches to the quantum theory the 
constraints---which are under much better control here than the evolution equations---are the 
fundamental objects \cite{WdW,Teitelboim}.  Indeed, in a Wheeler-DeWitt-type approach, a 
solution of the quantum constraints centered around some superposition of connected sums 
(\ref{z1}) would give a \emph{complete} description of the state.

Of course, time evolution must still be hidden somewhere in such a solution.   To extract
this behavior, we must address the notorious ``problem of time''  in quantum gravity \cite{Kuchar}.  
It has recently been proposed that solutions to the constraints contain all possible ``quantum
reference systems,'' with particular frames selected by the choice of gauge-fixing \cite{Hoehn}.
This suggests a new way to pose our question: does a typical solution of the Wheeler-DeWitt 
equation have a foamlike structure at the Planck scale in a generic quantum reference system?
More concretely, it may be possible to introduce a particular matter ``clock'' to
investigate the time evolution \cite{Brown,Marolf}.  For the short term, this might be easiest in a spherically
symmetric minisuperspace model based on \cite{Witt, Carlipz}.

It might be interesting to connect this approach to Hawking's four-dimensional Euclidean
spacetime foam \cite{Hawking}.  This would require a better understanding of the
four-dimensional evolution of our initial data.  But the ``necks'' in the connected sum (\ref{z1}) 
resemble throats of Schwarzschild black holes \cite{Burkhart,Burkhartb}, for which the Euclidean 
continuation is well understood, so progress may be possible.  It would also be worth looking
further at the $\Lambda<0$ case in light of the AdS/CFT correspondence.  Here, there has been 
interesting work on the question of which topologies contribute, although mainly in the context 
of black holes and lower dimensions \cite{Farey}.

\section{What this proposal does, and does not, do}

As early as 1957, Wheeler argued that
\begin{quote}
\dots it is essential to allow for fluctuations in the metric and gravitational interactions in any
proper treatment of the compensation problem---the problem of compensation of ``infinite'' energies that
is so central to the physics of fields and particles \cite{Wheelerb}.
\end{quote}
What I am proposing is a concrete realization of this vision.   Several previous attempts 
have been made to model spacetime foam---see, for instance, \cite{Hawking,Crane,Ng}---but only a 
few have addressed the cosmological constant problem \cite{Coleman,Kirillov,Garattini,Carlip}.  
The new ingredients here are the ability to construct a large new class of initial data and the 
crucial realization that time reversal invariance allows, and perhaps even requires, the expansion 
and shear to average to zero.

This proposal addresses the ``old'' cosmological constant problem, the problem of large
vacuum energy.  It does not tell us whether the observed accelerated expansion of the 
universe is caused by a small residual cosmological constant.  It has recently been argued   
that higher order correlations of vacuum fluctuations might generate a small cosmological 
constant \cite{Unruh,Unruhb,Santos}.  These would presumably show up here in higher 
correlations of the metric and extrinsic curvature, which appear in higher derivatives of 
averaged expansion and curvature. 

While this proposal offers a natural explanation for small macroscopic expansion and shear,
the requirement of small spatial curvature is less obvious.  It is 
certainly possible to choose data for which $\langle R\rangle$ is small, and there are hints that
this may be preferred by the gravitational partition function, but a better understanding is needed.
The answer may be dynamical.  In a standard closed FLRW cosmology, after all,
the spatial curvature is initially very high and decreases in time.  There is some evidence that
the same is true here: the second time derivative of the averaged curvature $\langle R\rangle$ 
can be calculated, and while the result depends on the
lapse function, most of the terms are negative definite.

The proposal also does not attempt to explain the emergence of a macroscopic arrow of time,
an important question but one that is probably logically independent. 
Nor have I shown that long wavelength disturbances sitting on top of Planck-scale spacetime 
foam will be described by classical general relativity.  This is the notorious ``averaging problem'' 
\cite{Clarkson,Green,Buchert}, the problem of how the nonlinearities of general relativity
interact with the process of taking averages.  Here, though, effective field 
theory arguments may help \cite{EFT}.   Nothing in this construction has broken
spatial diffeomorphism invariance, so the effective action should involve only terms
invariant under that symmetry.  This implies a Ho{\v r}ava-Lifshitz action \cite{Horava},
of which general relativity is a special case.  If, as I have argued, there is also
nothing ``preferred'' about the initial time slice, then time reparametrization invariance 
should also be a symmetry, in which case the large scale effective action should take the usual 
Einstein-Hilbert form.

So far, I have treated a quantum gravitational problem semiclassically, appealing to quantum
mechanics to generate Planck-scale structure but relying on classical general relativity to
describe constraints and evolution.   We might next consider coherent states centered on the 
configurations described here, and construct more general wave functions as superpositions.  But 
this would force us to confront some of the standard problems of quantum gravity: the metric and 
extrinsic curvature are not true observables, and to define an average we would have to figure out 
what ``at the same point'' means in  different components of the wave function.
 
Interesting technical questions remain as well.  The gluing construction I 
have used provides a large set of initial data, but it is not known just how much
of the total space of initial data is covered.   More generally, gluing is certainly not the \emph{only} way
to produce data with no arrow of time at the Planck scale, and a full understanding of the measure
such data is still lacking.   It would also be useful to further investigate higher order correlations, or, 
conversely, to see to what extent further restrictions (e.g., $\langle\mathcal{L}_n^3K\rangle=0$) 
limit the possible initial data.   

For all these limitations, though, this proposal suggests a simple and radical solution to a deep 
problem.  If a large cosmological constant is generated by vacuum fluctuations at the
Planck scale, then perhaps that is also the place to look for answers.  I have shown that at 
least in principle, hiding a Planck scale cosmological constant in Planck scale curvature fluctuations
is not only possible, but can be quite natural.  We may have simply been looking in the wrong 
place.

\begin{flushleft}
\large\bf Acknowledgments
\end{flushleft}

I would like to thank Andy Albrecht, Sean Carroll, Piotr Chrusciel, Markus Luty, Don Page, 
Daniel Pollack, Albert Schwarz, and Bob Wald for helpful discussions.  This work was supported 
in part by Department of Energy grant DE-FG02-91ER40674.


\newpage
\small

\begin{thebibliography}{99}
\bibitem{Holland} J.\ Holland and S.\ Hollands, ``A small cosmological constant due to 
non-perturbative quantum effects,'' Class.\ Quant.\ Grav.\ 31 (2014) 125006, arXiv:1305.5191.
\bibitem{Martin} J.\ Martin, ``Everything You Always Wanted To Know About The Cosmological 
Constant Problem (But Were Afraid To Ask),'' Comptes Rendus Physique 13 (2012) 566, arXiv:1205.3365.
\bibitem{EFT} C.~P.\ Burgess, ``Quantum gravity in everyday life: General relativity as an 
effective field theory,''  Living Rev.\ Rel.\ 7 (2004) 5, arXiv: gr-qc/0311082.
\bibitem{Carroll} S.~M.\ Carroll, ``The Cosmological constant,''  Living Rev.\ Rel.\ 4 (2001) 1,
arXiv:astro-ph/0004075.
\bibitem{Hobson} By M.~P.\ Hobson, G.~P.\ Efstathiou, and A.~N.\ Lasenby, \emph{General 
Relativity: An Introduction for Physicists} (Cambridge University Press, Cambridge, 2006), 
p.~187.
\bibitem{Weinberg} S.\ Weinberg, ``The Cosmological Constant Problem,''  Rev.\ Mod\ Phys.\
 61 (1989) 1.
\bibitem{Kaloper} N.\ Kaloper and A.\ Padilla, ``Sequestering the Standard Model Vacuum 
Energy,'' Phys.\ Rev.\ Lett.\ 112 (2014)  091304, arXiv:1309.6562.
\bibitem{Wheeler} J.~A.\ Wheeler, ``Geons,'' Phys.\ Rev.\ 97 (1955) 511.
\bibitem{Unruh} Q.\ Wang, Z.\ Zhu, and W.~G.\ Unruh, ``How the huge energy of quantum vacuum 
gravitates to drive the slow accelerating expansion of the universe,'' Phys.\ Rev.\ D95 (2017)
103504, arXiv:1703.00543.
\bibitem{Wald} R.~M.\ Wald, ``Asymptotic behavior of homogeneous cosmological models 
in the presence of a positive cosmological constant,'' Phys.\ Rev.\ D28 (1983) 2118.
\bibitem{Clarkson} C.\ Clarkson, G.\ Ellis, J.\ Larena, and O.\ Umeh, ``Does the growth of 
structure affect our dynamical models of the universe? The averaging, backreaction and
 fitting problems in cosmology,''  Rept.\ Prog.\ Phys. 74 (2011) 112901, arXiv:1109.2314.
 \bibitem{WdW} B.~S.\ DeWitt, ``Quantum Theory of Gravity 1. The Canonical Theory,''
Phys.\ Rev.\ 160 (1967)1113.
 \bibitem{Teitelboim} C.\ Teitelboim, ``Causality Versus Gauge Invariance in Quantum Gravity 
 and Supergravity,'' Phys.\ Rev.\ Lett.\ 50 (1983) 705.
\bibitem{Chrusciel} P.~T.\ Chrusciel, J.\ Isenberg, and D.\ Pollack, ``Gluing initial data sets 
for general relativity,'' Phys.\ Rev.\ Lett.\ 93 (2004) 081101, arXiv:gr-qc/0409047.
\bibitem{Chruscielb} P.~T.\ Chrusciel, J.\ Isenberg, and D.\ Pollack, ``Initial data engineering,''
Commun.\ Math.\ Phys.\ 257 (2005) 29, arXiv:gr-qc/0403066.
\bibitem{Milnor} J.\ Milnor, ``A Unique Decomposition Theorem for 3-Manifolds,'' Am.\ J.\ Math.\
84 (1962) 1.
\bibitem{top} D.\ Giulini, ``Properties of three manifolds for relativists,'' Int.\ J.\ Theor.\ Phys.\
 33 (1994) 913, arXiv:gr-qc/9308008.
 \bibitem{Witt} J.\ Morrow-Jones and D.~M.\ Witt. ``Inflationary initial data for generic spatial 
topology,'' Phys.\ Rev.\ D48 (1993) 2516.
\bibitem{Carlipz} S.\ Carlip, in preparation.
\bibitem{Burkhart} M.\ Burkhart, M.\ Lesourd, and D.\ Pollack, ``Null geodesic incompleteness
 of spacetimes with no CMC Cauchy surfaces,'' arXiv:1902.07411.
\bibitem{Unruhb} Q.\ Wang and W.~G.\ Unruh, ``Vacuum fluctuation, micro-cyclic `universes' 
and the cosmological constant problem,''  arXiv:1904.08599.
\bibitem{Buchertb} T.\ Buchert, ``A Cosmic equation of state for the inhomogeneous universe: 
Can a global far-from-equilibrium state explain dark energy?'' Class.\ Quant.\ Grav.\ 22 (2005) 
L113, arXiv:gr-qc/0507028.
\bibitem{Burkhartb} M.\ Burkhart and D.\ Pollack, ``Causal geodesic incompleteness of spacetimes
arising from IMP gluing,'' arXiv:1907.00295.
\bibitem{Pitelli} S.\ Carlip,  R.~A.\ Mosna, and J.~P.~M.\ Pitelli, ``Vacuum Fluctuations and the 
Small Scale Structure of Spacetime,'' Phys.\ Rev.\ Lett.\ 107 (2011) 021303, arXiv:1103.5993.
\bibitem{Green} S.~R.\ Green and R.~M.\ Wald, ``A new framework for analyzing the effects 
of small scale inhomogeneities in cosmology,'' Phys.\ Rev.\ D83 (2011) 084020, arXiv:1011.4920.
\bibitem{Buchert} T.\ Buchert, ``On average properties of inhomogeneous fluids in general 
relativity: Dust cosmologies,'' Gen.\ Rel.\ Grav.\ 32 (2000) 105, arXiv:gr-qc/9906015.
\bibitem{Torre}  C.~G.\ Torre, ``Gravitational Observables and Local Symmetries,''
Phys.\ Rev.\ D48 (1993) 2373, arXiv:gr-qc/9306030.
\bibitem{Kuchar} K.~V.\ Kucha{\v r}, ``Time and Interpretations of Quantum Gravity,'' 
Int.\ J.\ Mod.\ Phys.\ D20 (2011) 3.
\bibitem{Hoehn} P.~A.\ H{\"o}hn, Universe 5 (2019) 116, arXiv:1811.00611.
\bibitem{Brown} J.~D.\ Brown and K.~V.\ Kucha{\v r}, ``Dust as a standard of space and time 
in canonical quantum gravity,'' Phys.\ Rev.\ D51 (1995) 5600, arXiv:gr-qc/9409001.
\bibitem{Marolf} J.~D.\ Brown and D.\ Marolf, ``On relativistic material reference systems,''
Phys.\ Rev.\ D53 (1996) 1835, arXiv:gr-qc/9509026.
\bibitem{Hawking} S.~W.\ Hawking, ``Space-Time Foam,'' Nucl.\ Phys.\ B144 (1978) 349.
\bibitem{Farey} R.\ Dijkgraaf, J.\ Maldacena, G.\ Moore, and E.\ Verlinde, ``A Black Hole Farey Tail,''
arXiv:hep-th/0005003.
\bibitem{Wheelerb} J.~A.\ Wheeler, ``On the Nature of quantum geometrodynamics,''  Annals Phys.\
 2 (1957) 604, p. 610.
\bibitem{Crane} L.\ Crane and L.\ Smolin, ``Renormalizability of General Relativity on a Background 
of Space-time Foam, Nucl.\ Phys.\ B267 (1986) 714.
\bibitem{Ng} Y.~J.\ Ng, ``Space-time foam,''  Int.\ J.\ Mod.\ Phys.\ D11 (2002) 1585, 
arXiv:gr-qc/0201022.
\bibitem{Coleman} S.~R.\ Coleman, ``Why There Is Nothing Rather Than Something: A Theory 
of the Cosmological Constant,'' Nucl.\ Phys.\ B310 (1988) 643.
\bibitem{Kirillov} A.~A.\ Kirillov and E~P.\ Savelova, ``On the value of the cosmological constant 
in a gas of virtual wormholes,'' Grav.\ Cosmol.\ 19 (2013) 92.
\bibitem{Garattini} R.\ Garattini, ``A space-time foam approach to the cosmological constant 
and entropy,'' Int.\ J.\ Mod.\ Phys.\ D4 (2002) 635, arXiv:gr-qc/0003090.
\bibitem{Carlip} S.\ Carlip, ``Space-time foam and the cosmological constant,'' Phys.\ Rev.\ Lett.\ 
79 (1997) 4071, arXiv:gr-qc/9708026.
\bibitem{Santos} E.\ Santos, ``The cosmological constant problem or how the quantum vacuum 
drives the slow accelerating expansion of the universe,'' arXiv:1805.03018.
\bibitem{Horava} P.\ Horava, ``Quantum Gravity at a Lifshitz Point,'' Phys.\ Rev.\ D79 (2009) 084008,
arXiv:0901.3775.





\end{thebibliography}
\end{document}